\documentclass[aps,pre,preprint,floatfix]{revtex4}

\usepackage[normalem]{ulem}
\usepackage{latexsym}
\usepackage[usenames]{color}
\usepackage{amsthm}
\usepackage{hyperref}
\usepackage{braket}
\newcounter{para}

\usepackage{amsmath}    % need for subequations
\usepackage{amsfonts}  %note how statements can be commented out
\usepackage{amssymb}
\usepackage{graphicx}
\usepackage{slashed}
\usepackage{fouridx}
\usepackage{tikz}

\usetikzlibrary{calc}

\theoremstyle{plain} 
\theoremstyle{plain} 
\theoremstyle{plain}  
\theoremstyle{plain} 
\theoremstyle{plain}  
\theoremstyle{remark} 
\theoremstyle{plain} 
\theoremstyle{remark}

\newcommand\CROSS[1]{%
  \hbox{%
    \vbox{
      \hrule
      \kern2.5pt
      \hbox{$#1$\,\strut}
    }%
  \vrule
  }\mskip\thickmuskip
}

\tikzset{
solid node/.style={circle,draw,inner sep=1.5,fill=black},
hollow node/.style={circle,draw,inner sep=1.5}
}

\newlength{\arrowsize}  
\pgfarrowsdeclare{biggertip}{biggertip}{  
  \setlength{\arrowsize}{0.5pt}  
  \addtolength{\arrowsize}{.5\pgflinewidth}  
  \pgfarrowsrightextend{0}  
  \pgfarrowsleftextend{-5\arrowsize}  
}{  
  \setlength{\arrowsize}{0.5pt}  
  \addtolength{\arrowsize}{.5\pgflinewidth}  
  \pgfpathmoveto{\pgfpoint{-5\arrowsize}{4\arrowsize}}  
  \pgfpathlineto{\pgfpointorigin}  
  \pgfpathlineto{\pgfpoint{-5\arrowsize}{-4\arrowsize}}  
  \pgfusepathqstroke  
}

\begin{document}
\title{Universality of oscillatory instabilities in fluid mechanical systems}

\author{Vladimir Garc\'ia-Morales$^1$}
 \email{vladimir.garcia@uv.es}

\author{Shruti Tandon$^{2,3}$}
%\affiliation{$^1$Department of Aerospace Engineering, Indian Institute of Technology Madras, Chennai 600036, India}
\author{J\"urgen Kurths$^4$}
 \email{Juergen.Kurths@pik-potsdam.de}

\author{R. I. Sujith$^{2,3}$}
\email{sujith@iitm.ac.in}
\affiliation{$^1$Departament de F\'{\i}sica de la Terra i Termodin\`amica, Universitat de Val\`encia, E-46100 Burjassot, Spain}
\affiliation{$^2$Department of Aerospace Engineering, Indian Institute of Technology Madras, Chennai 600036, India}
\affiliation{$^3$Centre for Excellence for studying Critical Transitions in Complex Systems, Indian Institute of Technology Madras, Chennai 600036, India}
\affiliation{$^4$Potsdam Institute for Climate Impact Research, 14473 Potsdam, Germany}
\begin{abstract}
Oscillatory instability (OI) emerges amidst turbulent states in experiments in various turbulent fluid and thermo-fluid systems such as aero-acoustic, thermoacoustic and aeroelastic systems. For the time series of the relevant dynamic variable at the onset of the OI, universal scaling behavior have been discovered in experiments via the Hurst exponent and certain spectral measures. By means of a center manifold reduction, the spatiotemporal dynamics of these real systems can be mapped to a complex Ginzburg-Landau equation with a linear global coupling (GCGLE). In this work, we show that the GCGLE is able to capture the universal behavior of the route to OI, elucidating it as a transition from defect to phase turbulence mediated by the global coupling.
\end{abstract}
\maketitle

\section{Introduction}
Spatially extended ensembles of globally coupled nonlinear oscillators exhibit a wide variety of spatiotemporal patterns ranging from incoherence to uniform oscillations \cite{Mikhailov}. For example, in certain surface chemical reactions, global coupling is produced through mechanical interactions with the gas phase \cite{Kim, Veser}. Global coupling can lead to the emergence of synchronous behavior out of turbulence and the formation of a number of coherent structures \cite{Veser}.

Turbulent aero-acoustic \cite{Flandro}, thermoacoustic  \cite{Juniper}
and aeroelastic systems \cite{Hansen} can be viewed as paradigmatic spatially extended ensembles of globally coupled nonlinear oscillators. In all these systems, the behavior of an averaged dynamic variable reflects self-organization in the microscopic degrees of freedom leading to global periodic oscillations emerging out of turbulent states. The relevant dynamic variable is the acoustic pressure in aero-acoustic and thermoacoustic systems, and the strain experienced by the cantilever in aeroelastic systems. A transition to a state, termed \emph{oscillatory instability} (OI), occurs in the time series of this variable from low-amplitude aperiodic oscillations to high-amplitude periodic ones \cite{sujithbook, PavithranEPL}. As the control parameter of the turbulent flow (the Reynolds number $Re$) is varied in experiments, bursts of periodic oscillations begin to emerge amidst chaotic fluctuations. This alternating periodic and aperiodic behavior, \textcolor{black}{termed as intermittency,} has been reported in turbulent aero-acoustic \cite{nairacoustic}, aeroelastic \cite{aeroelasticSarkar} and thermoacoustic systems \cite{Nair}. The periodic content of the time series increases as the OI is approached \cite{PavithranEPL}.  

\textcolor{black}{Recently, experiments have established that the relevant variable of the OI exhibits power law scaling behavior via the Hurst exponent \cite{PavithranEPL} and a spectral measure (moment of the power spectrum) \cite{PavithranEPL} during the emergence of order. Strikingly, the power law scaling exponents are universally same across different turbulent fluid and thermo-fluid systems. Here, we propose a novel perspective where the relevant variable is viewed as} a spatial average of an order parameter obtained by reducing the original dynamics of the real fluid mechanical systems to an apropriate amplitude equation of a complex Ginzburg-Landau type. We show that, indeed, the cubic complex Ginzburg-Landau equation with a global linear coupling (GCGLE) is able to capture the emergence of OI along with the universal scaling behaviors found in experiments.

OIs emerge from turbulent states due to internal nonlinear interactions  \cite{EurJ}. For example, flame dynamics, acoustics and hydrodynamics are strongly coupled inside thermoacoustic systems, as found in gas turbines (for power-producing, aeronautical, and marine applications) \cite{Lieuwen} and rocket engines \cite{Fisher}. A positive feedback develops between the acoustic field and the heat release rate oscillations leading to high-amplitude self-sustained periodic oscillations that are catastrophic to these engineering systems. Similarly, the OI in aero-acoustic systems arises due to the coupling between the acoustic field and the vortex shedding in the flow within a confinement \cite{EurJ}. Such oscillatory dynamics has been studied in various aero-acoustic applications, such as driven cavities, jet noise, pipe whistling, and gas pipe networks \cite{Tonon}. Finally, the OI in aeroelastic systems occurs due to the interaction between the flow field and the structural elements of the system \cite{EurJ}. A classic example is the collapse of the Tacoma bridge in 1940 due to violent oscillations \cite{Larsen}.

In this work, we formulate a mathematical model (discussed in Sec. \ref{sec_model}) that is able to capture the universality observed in the transition to OI through experiments in fluid mechanical systems, reproducing both scaling laws of the amplitude of the dominant mode with the Hurst exponent and with the spectral measure (discussed in Sec. \ref{sec_results}). 

%\section{\label{sec_expt}Experimental findings}

\section{\label{sec_model}Model and computational details}

\textcolor{black}{We introduce a globally coupled Complex-Ginzburg Landau equation (GCGLE) to model the turbulent flow and nonlinear interactions in complex turbulent fluid systems.} Our model can be understood as arising from a center manifold reduction of the original dynamics (which is a vector space that for example in thermoacoustic systems, consists of combustion, acoustic and hydrodynamic subspaces that are interdependent due to nonlinear interactions) leading to the spatiotemporal dynamics of an oscillatory field. \textcolor{black}{The global coupling term captures the average strength of interactions between subsystems in a turbulent fluid and thermo-fluid system.}

The only free parameter of our theory is the global coupling strength $\gamma$. We assume that variations in the Reynolds number ($Re$) have an impact on $\gamma$ and on the parameters governing the local dynamics of the oscillators through their dependence on $\gamma$. Our model is a GCGLE \cite{Mikhailov, Kim, Veser} of the form
\begin{eqnarray}
\partial_{t}W &=& W+\left(1+ic_1(\gamma)\right)\partial_{x}^{2}W \nonumber \\
\qquad \qquad &&-\left(1+ic_2(\gamma)\right)\left|W\right|^2 W  \nonumber \\
\qquad \qquad &&+~\gamma(W-\left<W\right>) \label{GCGLE}
\end{eqnarray}
This equation describes the spatiotemporal evolution (space $x$, time $t$) of a complex order parameter $W(x,t)$ characterizing a 1D spatially extended array of diffusively and globally coupled nonlinear oscillators past a supercritical Hopf bifurcation. The term $\gamma(W-\left<W\right>)$ represents the deviation between the local oscillators and the global average dynamics $\left<W\right>$, where the brackets $\left<...\right>$ denote spatial averages. The parameter $\gamma$ is the strength of the global coupling and can be either desynchronizing ($\gamma<0$) or synchronizing ($\gamma>0$). For example, in thermoacoustic systems, it represents the coupling of the global acoustic field with the hydrodynamic and combustion subsystems in a turbulent thermoacoustic system. Further, in such systems, the underlying turbulent hydrodynamic and heat release rate fluctuations introduce diffusive coupling and nonlinearities which are respectively modeled by the terms $(1+ic_1)\partial_{x}^{2}W$ and $-(1+ic_2)\left|W\right|^2 W$ in Eq. (\ref{GCGLE}), respectively. The real-valued functions $c_{1}(\gamma)$ and $c_{2}(\gamma)$ can be obtained from the oscillatory dynamics of any specific system by means of a center manifold reduction \cite{Aranson, Kuramoto, contemphys}. They can be both expressed as a power series of $\gamma$ as
\begin{eqnarray}
c_1(\gamma)&=&\sum_{k=0}^{\infty} c_{1}^{(k)}\gamma^{k} \approx c_{1}^{(0)}+c_{1}^{(1)}\gamma \label{c1} \\
c_2(\gamma)&=&\sum_{k=0}^{\infty} c_{2}^{(k)}\gamma^{k} \approx c_{2}^{(0)}+c_{2}^{(1)}\gamma \label{c2}
\end{eqnarray}
The truncation of the above series to first order in $\gamma$ is justified by the fact that $\gamma$ is small. This warrants that Eq. (\ref{GCGLE}) above can be derived from a center manifold reduction of the original dynamics \cite{VGMKrisch}. Here, we shall consider $-0.01\le \gamma \le 0.0038$. The values of the parameters $c_{i}^{(j)}$, $i=1,2$, $j=0,1$ are fixed in all simulations to  $c_1^{(0)}=8.25$, $c_1^{(1)}=-798.95$, $c_2^{(0)}=-0.75$, $c_2^{(1)}=51.05$. The exhaustive search leading to these values is guided by the fact that when $\gamma=0$, the system is to be found in a defect turbulent state (combustion noise) and, therefore, $1+c_1^{(0)}c_2^{(0)} <<0$, i.e. the system is found deep in the Benjamin-Feir (BF) unstable regime. As $\gamma >0$, the inequality $1+c_1(\gamma)c_2(\gamma) <0$ is not so strong because parameters $c_1$ and $c_2$ governing the local dynamics are driven by the global coupling to a situation where the uniform oscillation is gradually stabilized and, therefore, closer to the BF line.

Let $p(t)$ denote the pressure fluctuations in aero-acoustic and thermoacoustic systems or the strain experienced by the cantilever in aeroelastic systems. We model $p(t)$, regardless of the specific system under consideration, as
\begin{equation}
p(t)=\frac{\lambda}{L}\int_{0}^{L}\text{Re}W(x,t)\text{d} x=\lambda \left< \text{Re}W\right> \label{pt}
\end{equation}
where $\lambda$ is a constant. $L$ is the (dimensionless) system size and $\text{Re}W(x,t)$ denotes the real part of  $W(x,t)$. We fix $L=100$ and $\lambda=0.83$ kPa in all simulations.  The value of $L$ is chosen so that the system is macroscopic and the spatial average is robust. The value of $\lambda$ is obtained from fitting the amplitude of $p(t)$ when all oscillators are synchronized \cite{PavithranEPL}. In brief, since $\text{Re} W$ directly models the experimental spatiotemporal dynamics, the relevant dynamical variable $p(t)$ is merely proportional to the spatial average $\left<\text{Re} W\right>$. (The imaginary part of the order parameter, $\text{Im} W(x,t)$, corresponds to the Hilbert transform of the experimental spatiotemporal dynamics \cite{synch}.)

By applying periodic boundary conditions, the general solution of Eq. (\ref{GCGLE}) can be written as
\begin{equation}
W(x,t)=\sum_{n=-\infty}^{\infty}W_{n}(t)\text{e}^{i2\pi n x/L} \label{gensol}
\end{equation}
By replacing this general solution in Eq. (\ref{GCGLE}) we find for the coefficients $W_n(t)$ that
\begin{eqnarray}
\dot{W}_{n}&=&\left[1+\gamma(1-\delta_{n0})-(1+ic_1)\left(\frac{2\pi n}{L} \right)^2 \right]W_{n} \nonumber \\
&&-(1+ic_2)\sum_{j-k+l=n}W_{j}\widetilde{W}_{k}W_{l} \label{FCGLE}
\end{eqnarray}
where $\delta_{n0}=1$ if $n=0$ and $\delta_{n0}=0$ otherwise. Here the tilde denotes complex conjugation and the sum runs over all integer values for $j$, $k$ and $l$ such that $j-k+l=n$. The sum is truncated for a number $N=512$ of Fourier modes and Eq. (\ref{FCGLE}) is numerically solved by means of the ETD2K exponential time-stepping algorithm of Mathews and Cox \cite{Cox}. The solution for $W(x,t)$ is then obtained from Eq. (\ref{gensol})), and $p(t)$ is calculated from Eq. (\ref{pt}).

\section{\label{sec_results}Results}

\textcolor{black}{Oscillatory instability emerges when a fluctuation drives the system to self-sustained dynamics due to the self-organised feedback interactions within the system. In experiments, as the flow control parameter ($Re$) is increased, the strength of nonlinear interactions increases and governs the dynamics. To model such increase in nonlinear interactions, here, we increase the global coupling parameter of the GCGLE, changing it from $\gamma<0$ to $>0$, while a local pulse perturbation simulates the initial fluctuation. Clearly, the GCGLE gives rise to intermittent periodic and self-sustained periodic oscillations for  $\gamma>0$ and chaotic dynamics for $\gamma<0$.}

\begin{figure*}[t]
\begin{center}
\includegraphics[width=0.8 \textwidth]{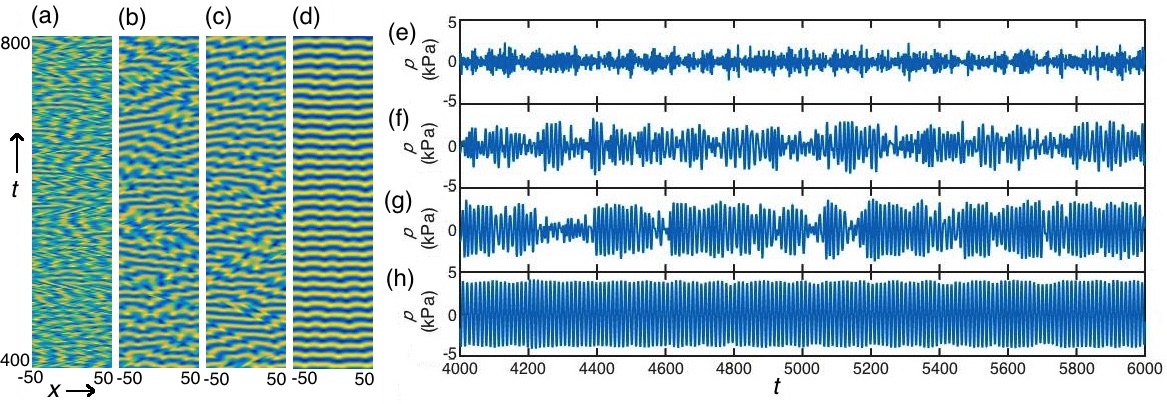}
\caption{\scriptsize{Spatiotemporal evolution of $\text{Re}W(x,t)$, where $x \in [-50, 50]$ and $t\in [400,800]$ obtained from the numerical integration of Eq. (\ref{GCGLE}) with a localized pulse as initial condition for $\gamma=-0.01$ (a), $\gamma=0$ (b), $\gamma=0.001$ (c), $\gamma=0.0038$ (d). The respective time series for $p(t)$ obtained from Eq. (\ref{pt}) are shown in panels (e)-(h) for longer time spans, with $t\in [4000, 6000]$.
}} \label{spatio}
\end{center}
\end{figure*}

In Fig. \ref{spatio}(a)-(d), the spatiotemporal evolution of $\text{Re}W(x,t)$, for a localized pulse as initial condition, is shown for increasing values of $\gamma$: we have $\gamma=-0.01$ (a), $\gamma=0$ (b), $\gamma=0.001$ (c), $\gamma=0.0038$ (d). The respective time series for $p(t)$ are shown in panels (e)-(h) for longer time spans.  For $\gamma=0$, from Eqs. (\ref{c1}) and (\ref{c2}) we have $c_1=c_1^{(0)}=8.25$ and  $c_2=c_2^{(0)}=-0.75$. Since $c_1c_2=-6.1875 < -1$, when $\gamma=0$ the system is found in a state deep beyond the BF line, where the uniform oscillation is unstable to long wavelength perturbations. Indeed, the system exhibits \emph{defect turbulence}: the presence of defects, i.e. those $(x,t)$ pairs where $|W(x,t)|=0$, causes discontinuities in the lines of constant phase yielding an irregular spatiotemporal pattern where the oscillators are incoherent, see Fig. \ref{spatio}(b). Now, if $\gamma <0$, as in Fig. \ref{spatio}(a) where $\gamma=-0.01$, the density of defects is even higher because the coupling is de-synchronizing. On the contrary, if the global coupling is positive, as in Fig. \ref{spatio}(c) and (d), defect turbulence tends to be suppressed. The density of defects diminishes until they are fully suppressed and only slight phase modulations remain, the lines of constant phase being continuous everywhere and the oscillators being synchronous.

The impact of this spatiotemporal behavior on $p(t)$ now becomes qualitatively clear. In Fig. \ref{spatio}(e)-(h), $p(t)$ is shown for different values of $\gamma$. We note that $0\le |p(t)| \le p_{max}$, where the upper bound $p_{max}=\lambda N /L=4267$ Pa is reached only when all oscillators are synchronous, all constructively contributing to the spatial average in Eq. (\ref{pt}). The lower bound $|p|=0$ is obtained in the extreme case in which the density of defects is so high that the oscillators are fully incoherent so that their spatial average vanishes. This situation is approached in Fig. \ref{spatio} (e), where $\gamma=-0.01$. 

As $\gamma$ is increased to $\gamma=0$ in Fig \ref{spatio} (f) and $\gamma=0.001$ in Fig \ref{spatio} (g), the time series tends to become intermittent owing to a gradual elimination of the defects in the spatiotemporal dynamics of oscillators. Hence, there are epochs when a majority of the oscillators are synchronized, while at other epochs they are desynchronized. This leads to bursts of periodic behavior interspersed between intervals of low-amplitude aperiodicity. During the epochs of aperiodic dynamics, the presence of defects makes the oscillators contribute destructively to $p(t)$. 

Finally, for $\gamma=0.0038$ as in Fig \ref{spatio} (h), the upper bound $|p|=p_{max}$ is attained periodically, because defects have been fully suppressed by the positive global coupling leading to the complete synchronization of all oscillators. Thus, all oscillators contribute constructively to the integrand in Eq. (\ref{pt}) when at maximum amplitude. In this way, our theory elucidates the emergence of OI. \textcolor{black}{We conclude that, the transition from low-amplitude aperiodic oscillations to large-amplitude periodic ones via the route of intermittency in $p(t)$ is essentially a transition from a state of defect to phase turbulence in the GCGLE as a consequence of increase in linear global coupling.}

The time series of $p(t)$ in Fig. \ref{spatio}(e)-(h) derived from the mean-field behavior of GCGLE accurately replicate the dynamics of realistic time series data obtained from experiments in thermoacoustic, aero-acoustic and aeroelastic systems. To substantiate this statement, we now calculate the Hurst exponent $H$ of $p(t)$ for an ensemble of 3000 simulations of our model at different $\gamma$ values in the interval $0\le \gamma \le 0.0038$. Generally, $H$ has values between 0 and 1 for time series (i.e., a fractal dimension between 1 and 2), and provides a measure of persistence  \cite{Feder}. An antipersistent signal has $H < 0.5$, in which a high value of the signal is most likely followed by a low value, whereas for a persistent signal $H > 0.5$. Note, $H = 0.5$ corresponds to an uncorrelated random process \cite{PavithranEPL}.   

\begin{figure}  
\begin{center}
\includegraphics[width=0.52\textwidth]{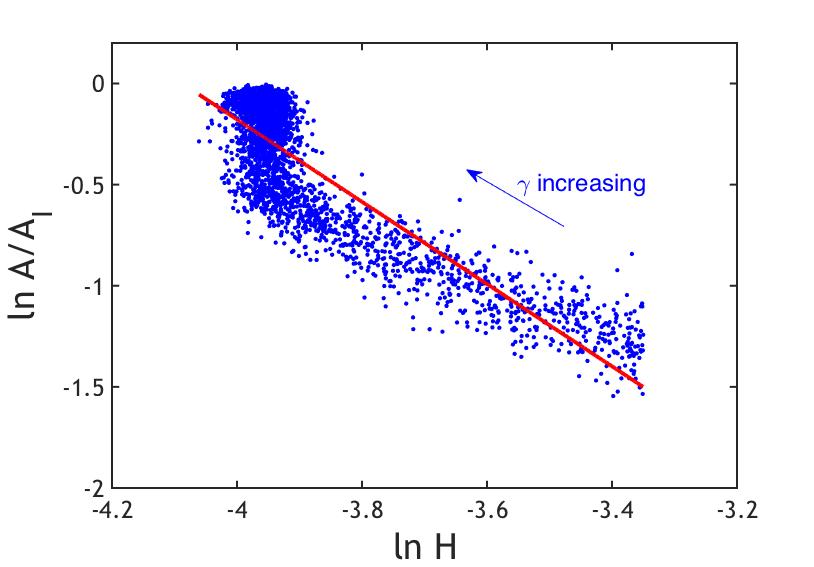}
\caption{\scriptsize{Normalized amplitude of the dominant Fourier mode plotted as a function of the Hurst exponent for 3000 time series obtained from Eqs. (\ref{GCGLE}) to (\ref{c2}). The scaling relationship $A/A_{I} \propto H^{-2.0\pm 0.1}$ is obtained (red line).
}} \label{hurst}
\end{center}
\end{figure} 

The Hurst exponent $H$ is used to predict the onset of OI in experimental systems. $H$ quantifies the scaling of the root mean squared (rms) of the standard deviation of fluctuations with the scale size or the time interval considered for obtaining the fluctuations. Recently, it was reported that the amplitude $A$ of the dominant mode of oscillations increases and follows a universal scaling law $A \propto H^{-2.0\pm 0.2}$ \cite{PavithranEPL}. A spectral measure $\mu$, which quantifies the sharpening of peaks in the power spectrum as the system dynamics approaches OI, has also been introduced and another universal scaling law $A\propto \mu^{-0.66\pm 0.10}$ \cite{PavithranSCIREP, Pavithesis} has been discovered. In both cases, the scaling exponents are invariant across aero-acoustic, thermoacoustic and aeroelastic systems. 

In Figure \ref{hurst}, the logarithm of the normalized amplitude of the dominant Fourier mode $A/A_{I}$ is plotted as a function of the logarithm of $H$ for the ensemble of time series obtained with our model. As $\gamma$ is increased, the Hurst exponent decreases and the amplitude of the dominant mode increases until the dominant mode coincides with the mode of the uniform synchronous oscillation. We find that the scaling relationship obtained from our model is $A/A_{I}\propto H^{-2.0\pm 0.1}$, which is in excellent agreement with the experimental results from fluid mechanical systems \cite{PavithranEPL}. 

\begin{figure}
\begin{center}
\includegraphics[width=0.52\textwidth]{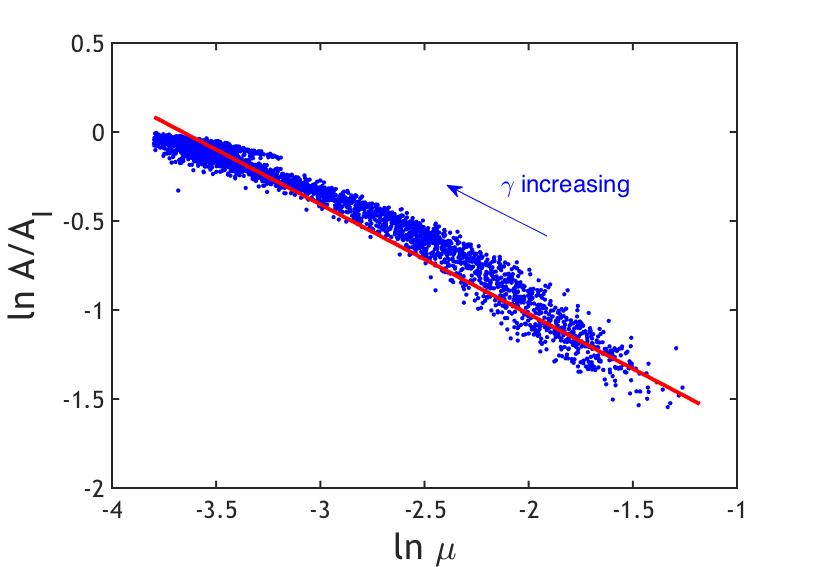}
\caption{\scriptsize{Normalized amplitude of the dominant Fourier mode plotted as a function of the spectral measure $\mu$ for the 3000 time series obtained from Eqs. (\ref{GCGLE}) to (\ref{c2}) and used in Fig. \ref{hurst}. The scaling relationship $A/A_{I}\propto \mu^{-0.61\pm 0.04}$ is obtained (red line). }} \label{measure}
\end{center}
\end{figure}

Furthermore, a transition of power spectra from a broad peak to a sharp one is obtained as the onset of the OI is approached \cite{PavithranSCIREP}. Spectral measures which quantify the sharpening of peaks in the power spectrum have been introduced. They are calculated as the product of different moments of the normalized power spectrum raised to integer powers, and they follow inverse power law relations with the corresponding peak power \cite{PavithranSCIREP}. The most prominent spectral measure $\mu$ is defined as
\begin{equation}
\mu:=\left[\int_{-\delta F}^{\delta F}\frac{P(F)}{P_0}\left|\frac{F}{f_0} \right|^2\text{d}F\right] \times \left[\int_{-\delta F}^{\delta F}\frac{P(F)}{P_0}\text{d}F\right] 
\end{equation}
Here $P(F)$ is the power corresponding to the modified frequency $F = f -f_0$, where $f$ indicates the frequency of oscillations, $f_0$ is the frequency corresponding to the dominant peak in the power spectrum, and $P_0:= P(f_0)$. A universal scaling relationship relating the normalized amplitude $A/A_I$ of the dominant mode of oscillations and the spectral measure $\mu$ has been obtained from experiments as $A\propto \mu^{-0.66\pm 0.10}$ \cite{PavithranSCIREP}. In Fig. \ref{measure} the normalized amplitude of the dominant Fourier mode is plotted as a function of the spectral measure $\mu$ for the ensemble of time series obtained from the GCGLE model (Eqs. \ref{GCGLE} to \ref{c2}). The scaling relationship $A/A_{I} \propto \mu^{-0.61\pm 0.04}$ is retrieved, also in excellent agreement with the scaling relationship obtained from experiments.

\begin{figure}
\begin{center}
\includegraphics[width=0.52\textwidth]{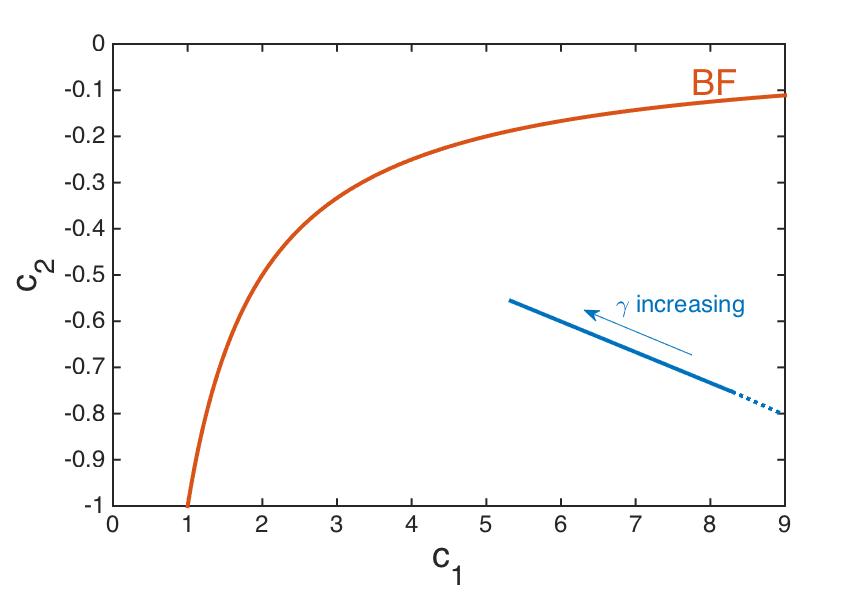}
\caption{\scriptsize{Location in the $c_2-c_1$ plane of the parameter values obtained from Eqs. (\ref{c1}) and Eqs. (\ref{c2}) as $\gamma$ is varied from 0 to 0.0038 (continuous blue line) and for $\gamma$ negative (dotted line). Shown also is the Benjamin-Feir (BF) line.}} \label{c1c2}
\end{center}
\end{figure}

Figure \ref{c1c2} shows the location of the $(c_1,c_2)$ pairs obtained from Eqs. (\ref{c1}) and (\ref{c2}) in the $c_2-c_1$ plane as $\gamma$ is varied. The continuous blue line shows the $(c_1,c_2)$ pairs when $\gamma$ is varied between 0 to 0.0038 and by a dotted line for $\gamma<0$. All $(c_1,c_2)$ pairs fall in a region in the plane where the uniform oscillation is unstable when the complex Ginzburg-Landau equation without global coupling is considered. The effect of a positive global coupling in our model is twofold: 1) it tends to suppress turbulence and 2) it displaces the values of $c_1$ and $c_2$ closer to the BF line (also shown in Fig. \ref{c1c2}). The latter also has a stabilizing effect, since it removes unstable Fourier modes out of the turbulent patterns. These two effects are directly responsible for the behavior observed in the spectral measure in Fig. \ref{measure} as the global coupling is increased. 

The fixed parameter values entering Eqs. (\ref{c1}) and (\ref{c2}) have been found through an extensive and systematic search of the parameter space. Although spatiotemporal chaos and intermittency exist outside the region of turbulence bounded by the BF line \cite{Aranson}, computational evidence suggests that the scaling relationships in Figs. \ref{hurst} and \ref{measure} are nowhere else to be found in the parameter plane.

\section{\label{sec_conclude}Conclusion}

In this work, we propose a globally coupled complex Ginzburg Landau equation (GCGLE) for \textcolor{black}{modeling the transition from chaotic to periodic temporal dynamics} in thermoacoustic, aero-acoustic and aeroelastic systems as found in experiments. The model arises from a center manifold reduction of the complex oscillatory spatiotemporal dynamics exhibited by many \textcolor{black}{turbulent fluid and thermo-fluid} systems close to an oscillatory instability which involves nonlinear interactions between \textcolor{black}{hydrodynamic and acoustic modes and combustion dynamics or elasticity fields}. Our model leads to an interpretation of the \textcolor{black}{emergence of} OI as a transition from defect to phase turbulence with the global coupling becomes more positive. Importantly, the universal scaling relationships discovered in experiments are accurately reproduced by our theory, and, therefore, the model is able to realistically replicate time series obtained from a wide variety of \textcolor{black}{turbulent} fluid mechanical systems.

\section*{Acknowledgements:}
 
R.I.S. acknowledges the funding from J. C. Bose Fellowship (No. JCB/2018/000034/SSC) and the IoE initiative (SB/2021/0845/AE/MHRD/002696). S.T. acknowledges the support from Prime Minister Research Fellowship, Govt. of India.

%%%%%%%%%%%%%%%%%%%%%%%%%%%%%%%%%%%%%%%%%%%%%%%%%%%%%%%%%%%%%%%%%%%%%%%%%%%%%%%%%%%%%%%%


\section*{References:}

\begin{thebibliography}{9}

\bibitem{Mikhailov}
A. S Mikhailov and K. Showalter, Phys. Rep. \textbf{425}, 79 (2006).
\bibitem{Kim}
M. Kim, M. Bertram, M. Pollmann, A. von Oertzen, A. S. Mikhailov, H. H. Rotermund, and G. Ertl, Science \textbf{292}, 1357 (2001); R. Imbihl, Prog. Surf. Sci. \textbf{44}, 185 (1993).
\bibitem{Veser}
G. Veser, F. Mertens, A. S. Mikhailov and R. Imbihl, Phys. Rev. Lett. \textbf{71}, 935 (1993); F. Mertens, R. Imbihl, and A. Mikhailov, J. Chem. Phys. \textbf{99}, 8668 (1993); \textbf{101}, 9903 (1994); M. Falcke, H. Engel, and M. Neufeld, Phys. Rev. E \textbf{52}, 763 (1995); M. Falcke and H. Engel, J. Chem. Phys. \textbf{101}, 6255 (1994); D. Battogtokh and A. Mikhailov, Physica D \textbf{90}, 84 (1996).
\bibitem{Flandro}
G. A. Flandro and J. Majdalani, AIAA J., \textbf{41}, 485 (2003),
\bibitem{Juniper}
M. P. Juniper and R. I. Sujith, Annu. Rev. Fluid Mech.,
\textbf{50}, 661 (2018).
\bibitem{Hansen}
H. M. Hansen, Wind Energy, \textbf{10}, 551 (2007).
\bibitem{sujithbook}
R. I. Sujith and S. A. Pawar, \emph{Thermoacoustic instability: A complex systems perspective}. (Springer, New York, 2021).
\bibitem{nairacoustic}
V. Nair and R. I. Sujith, Int. J. aero-acoustics., \textbf{15}, 312 (2016).
\bibitem{aeroelasticSarkar}
J. Venkatramani, V. Nair, R. I. Sujith, S. Gupta and S. Sarkar, J. Sound Vib. \textbf{386}, 390 (2017).
\bibitem{Nair}
V. Nair, G. Thampi and R. I. Sujith, J. Fluid Mech. \textbf{756}, 470 (2014).
\bibitem{PavithranEPL}
I. Pavithran, V. R. Unni, A. J. Varghese, R. I. Sujith, A. Saha, N. Marwan, and J. Kurths, Europhys. Lett., \textbf{129}, 24004 (2020).
\bibitem{EurJ}
I. Pavithran, V. R. Unni, and R. I. Sujith, Eur. Phys. J. Spec. Top. \textbf{230}, 3411 (2021)
\bibitem{Lieuwen}
T. C. Lieuwen and V. Yang, \emph{Combustion Instabilities in Gas Turbine Engines: Operational Experience, Fundamental Mechanisms, and Modelling}. (American Institute of Aeronautics and Astronautics, Reston VA, 2005). 
%Prog. Astronaut. Aeronaut. \textbf{210} (2005).
\bibitem{Fisher}
S. C. Fisher and S. A. Rahman, \emph{Remembering the giants: Apollo rocket propulsion development}, NASA/SP-2009-4545 (2009)
\bibitem{Tonon}
D. Tonon, A. Hirschberg, J. Golliard, S. Ziada, Int. J. Aeroacoust. \textbf{10}, 201 (2011)
\bibitem{Larsen}
A. Larsen, J.H. Walther, J. Wind Eng. Ind. Aerodyn. \textbf{67}, 253 (1997)
\bibitem{PavithranSCIREP}
I. Pavithran, V. R. Unni, A. J. Varghese, D. Premraj, R. I. Sujith, C.  Vijayan, A. Saha, N. Marwan, and J. Kurths, Sci. Rep. \textbf{10}, 17405 (2020).
\bibitem{Pavithesis}
I. Pavithran, V. R. Unni, A. Saha, A. J. Varghese,
R. I. Sujith, N. Marwan, J. Kurths, J. Eng. Gas Turbines Power. \textbf{143}, 121005 (2021).
\bibitem{Kuramoto}
Y. Kuramoto, \emph{Chemical Oscillations, Waves and Turbulence}  (Springer-Verlag, Berlin, 1984).
\bibitem{Aranson}
I. S. Aranson and L. Kramer, Rev. Mod. Phys. \textbf{74}, 99 (2002).
\bibitem{contemphys}
V. Garc\'{\i}a-Morales and K. Krischer, Contemp. Phys., \textbf{53}, 79 (2012).
\bibitem{VGMKrisch}
V. Garc\'{\i}a-Morales and K. Krischer, Phys. Rev. E \textbf{78}, 057201 (2008).
 \bibitem{synch}
A. Pikovsky, M. Rosenblum, J. Kurths, \emph{Synchronization: A Universal Concept in Nonlinear Sciences}  (Cambridge University Press, Cambridge UK, 2001).
\bibitem{Cox}
S. M. Cox, P. C. Matthews, J. Comput. Phys. \textbf{176}, 430 (2002).
\bibitem{Feder}
J. Feder, \emph{Fractals}. (Plenum Press, New York, 1988).


\end{thebibliography}
\end{document}